\documentclass[twocolumn,showpacs,preprintnumbers,superscriptaddress,
amsmath,amssymb]{revtex4}

\usepackage{epsfig}
\usepackage{graphicx}
\usepackage{dcolumn}
\usepackage{bm}
\usepackage{times}
\usepackage{color}
\usepackage{multirow}
\usepackage{cases}

\begin{document}


\title{Playing the role of weak clique property in link prediction: A friend recommendation model}

\author{Chuang Ma}
\affiliation{School of Mathematical Science, Anhui University, Hefei 230601, China}

\author{Tao Zhou}
\affiliation{Web Sciences Center, University of Electronic
Science and Technology of China, Chengdu 610054, China}


\author{Hai-Feng Zhang}\email{haifengzhang1978@gmail.com}
\affiliation{School of Mathematical Science, Anhui University, Hefei 230601, China}

\affiliation{
Key Laboratory of Computer Network and Information Integration (Southeast University), Ministry of Education, 211189, P. R. China
}
\affiliation{
Center of Information Support \&Assurance Technology, Anhui University, Hefei  230601,  China
}

\date{\today}

\begin{abstract}
An important fact in studying the link prediction is that the structural properties of networks have significant impacts on the performance of algorithms. Therefore, how to improve the performance of link prediction with the aid of structural properties of networks is an essential problem. By analyzing many real networks, we find a common structure property:  nodes are preferentially linked to the nodes with the weak clique structure (abbreviated as PWCS to simplify descriptions). Based on this PWCS phenomenon, we propose a local friend recommendation (FR) index to facilitate link prediction. Our experiments show that the performance of FR index is generally better than some famous local similarity indices, such as Common Neighbor (CN) index, Adamic-Adar (AA) index and Resource Allocation (RA) index. We then explain why PWCS can give rise to the better performance of FR index in link prediction. Finally, a mixed friend recommendation index (labelled MFR) is proposed by utilizing the PWCS phenomenon, which further improves the accuracy of link prediction.
\end{abstract}


\maketitle

\section{Introduction} \label{sec:intro}

The research of link prediction mainly focuses on forecasting potential relations between nonadjacent nodes, including the prediction of the unknown links or the further nodes~\cite{getoor2005link}. Since the wide range of applications of link prediction, such as recommending friends in online social networks~\cite{scellato2011exploiting}, exploring protein-to-protein interactions~\cite{cannistraci2013link}, reconstructing airline network~\cite{guimera2009missing}, and boosting e-commerce scales, which has attracted much attention recently~\cite{lu2011link,lu2015toward,lu2009similarity}. The probabilistic model and machine learning were mainly introduced in link prediction.  The notion of probabilistic link prediction and path analysis using Markov chains method were first proposed and evaluated in Ref.~\cite{sarukkai2000link}, and then Markov chains method was further studied in adaptive web sites~\cite{zhu2002using}; In Ref.~\cite{popescul2003statistical}, Popescul \emph{et~al.} studied the application of statistical relational learning to link prediction in the domain of scientific literature citations.

However, the mentioned methods for link prediction were mainly based on attributes of nodes. It is known that the structure of the network is easier to be obtained than the attributes of nodes, as a result, the network-structure-based link prediction has attracted increasing attention. Along this line, Liben-Nowell \emph{et~al.} developed approaches to link prediction based on measures for analyzing the ``proximity'' of nodes in a network~\cite{liben2007link}.  Since hierarchical structure commonly exists in the food webs, biochemical networks, social networks and so forth, a link prediction method based on the knowledge of hierarchical structure was investigated in Ref.~\cite{clauset2008hierarchical}, and they found that such a method can provide an accurate performance. Zhou \emph{et~al.} proposed a local similarity index---Resource Allocation (RA) index to predict the missing links, and their findings indicate that RA index has the best performance of link prediction~\cite{zhou2009predicting}. Given that many networks are sparse and very huge, so Liu\emph{ et~al.} presented a local random walk method to solve the problem of missing link prediction, and which can give competitively good prediction or even better prediction than other random-walk-based methods while has a lower computational complexity~\cite{liu2010link}. In view of the local community features in many networks, Cannistraci\emph{ et~al.} proposed an efficient computational framework called local community paradigm to calculate the link similarity between pairs of nodes~\cite{cannistraci2013link}. Liu \emph{et~al.} designed a
parameter-free local blocking predictor to detect missing links in given networks via local link density calculations, which performs better than the traditional local indices with the same time complexity~\cite{liu2015local}.

Since the structural properties of networks have significant effects on the performance of algorithms in link predictions, there are some literatures have proposed some methods by making use of the structural properties of networks. Such as the algorithms by playing the roles of hierarchical structure~\cite{clauset2008hierarchical}, clustering~\cite{feng2012link}, weak ties~\cite{lu2011link} and local community paradigm~\cite{cannistraci2013link}. However, current advances in including structural properties into link prediction are still not enough. In this paper, by investigating the local structural properties in many real networks, we find a common phenomenon: nodes are preferentially linked to the nodes with weak clique structure (PWCS). Then based on the observed phenomenon, a friend recommendation (FR) index is proposed. In this method, when a node $j$ introduces one of his friends to a node $i$, he will not introduce the nodes who are also the neighbors of node $i$.  Our results show that the performance of FR index is significantly better than CN, AA and RA indices since FR index can make good use of PWCS in networks. At last, to further play the role of PWCS, we define a mixed friend recommendation (MFR) method, which can better improve the accuracy in link prediction.
%

\section{ Preliminaries} \label{sec:preliminaries}

Considering an undirected and
unweighed network $G(V, E)$, where $V$ is the set of nodes and $E$
is the set of links. The multiple links and self-connections
are not allowed.  For a network with size $N$, the universal set of all possible links, denoted by $U$, containing of $ \frac{N(N-1)}{2}$ pairs of links. For each pair of nodes, $x, y\in V$, we assign a score, $S_{xy}$, according to a defined similarity measure.
Higher score means higher similarity between $x$ and $y$, and
vice versa. Since G is undirected, the score is supposed to
be symmetry, that is $S_{xy} = S_{yx}$. All the nonexistent links are
sorted in a descending order according to their scores, and
the links at the top are most likely to exist~\cite{zhou2009predicting,liu2010link}. To test the prediction accuracy of each index, we
adopt the approach used in Ref.~\cite{zhou2009predicting}. The link set $E$ is randomly divided into two sets  $E=E^T \cup E^P$ with
$E^T \cap E^P=\emptyset$. Where set $E^T$ is the training set and is supposed to be
known information, and $E^P$ is the testing set for the purpose of
testing and no information therein is allowed to be used for
prediction. As in previous literatures, the training set $E^T$ always contains 90\% of links in this work,
and the remaining 10\% of links constitute the testing set.
Two standard metrics are used to quantify the accuracy of prediction algorithms: area under the receiver operating
characteristic curve (AUC) and Precision~\cite{lu2011link}.

Area under curve (AUC)  can be
interpreted as the probability that a randomly chosen missing
link (a link in $E^P$) is given a higher score than a randomly
chosen nonexistent link (a link in $U-E^P$). When implementing, among $n$
independent comparisons, if there are $n'$ times the missing
link has a higher score and $n''$ times they are the same score,
AUC can be read as follow~\cite{lu2011link}:
\begin{eqnarray}\label{1}
AUC=\frac{n'+0.5n''}{n}.
\end{eqnarray}

If all the scores generated from independent
and identical distribution, the accuracy should be about 0.5.
Therefore, the degree to which the accuracy exceeds 0.5
indicates how much the algorithm performs better than pure
chance.

Precision is the ratio
of the number of missing links predicted correctly within
those top-L ranked links to $L$, and $L=100$ in this paper. If $m$ links are correctly predicted, then Precision can be calculated as~\cite{lu2011link}:
\begin{eqnarray}\label{2}
Percision=\frac{m}{L}.
\end{eqnarray}

We mainly compare three local similarity indices for link prediction, including (1) Common Neighbors(CN)~\cite{newman2001clustering}; (2) Adamic-Adar (AA) index~\cite{adamic2003friends}; (3) Resource Allocation (RA) index~\cite{zhou2009predicting}. Among which, CN index is the simplest index.  AA index and RA index have the similar form, and they both depress the contribution of the high-degree common neighbors, however, Zhou \emph{et~al.} have shown that the performance of RA index is generally better than AA index.

Let $\Gamma(i)$ be the neighbor set of node $i$, $|.|$ be the cardinality of the set, and $k(i)$ be the degree of node $i$. Then CN index, AA index and RA index are defined as

\textbf{CN index}
\begin{eqnarray}\label{3}
S^{CN}_{ij}=|\Gamma(i)\cap\Gamma(j)|,
\end{eqnarray}

\textbf{AA index}

\begin{eqnarray}\label{4}
S^{AA}_{ij}=\sum_{\Gamma(i)\cap\Gamma(j)}\frac{1}{\lg(k(l))},
\end{eqnarray}

\textbf{RA index}

\begin{eqnarray}\label{5}
S^{RA}_{ij}=\sum_{\Gamma(i)\cap\Gamma(j)}\frac{1}{k(l)},
\end{eqnarray}
respectively.

\section{ Data Set} \label{sec:data}
In this paper, we choose twelve representative networks drawn
from disparate fields: including: (1) C. elegans-The neural network of the nematode worm C. elegans~\cite{watts1998collective}; (2) NS-A coauthorship network of scientists working on network theory and experiment~\cite{von2002comparative}; (3) FWEW-A 66 component budget of the carbon exchanges occurring during the wet and dry seasons in the graminoid ecosystem of South Florid~\cite{ulanowicz1998network}; (4) FWFW-A food web in Florida Bay during the rainy season~\cite{ulanowicz1998network}; (5) USAir-The US Air transportation system~\cite{link}; (6) Jazz-A collaboration network of jazz musicians~\cite{gleiser2003community}; (7) TAP-yeast protein-protein binding network generated by tandem affinity purification experiments~\cite{gavin2002functional}; (8) Power-An electrical power grid of the western US~\cite{watts1998collective}; (9) Metabolic-A metabolic network of C.elegans~\cite{duch2005community}; (10) Yeast-A protein-protein interaction network in budding yeast~\cite{bu2003topological}; (11) Router-A symmetrized snapshot of the structure of the Internet at the level of autonomous systems~\cite{spring2004measuring}; (12) PB-A network of the US political blogs~\cite{reese2007mapping}. Topological features of these networks are summarized in Tab.~\ref{table1}.

\begin{table}
\centering
\caption {The basic topological features of twelve example networks.$N$ and $M$ are the total numbers of nodes and links,
respectively.  $C$ and $r$ are clustering coefficient and assortative coefficient, respectively. $H$ is the
degree heterogeneity, defined as $H=\frac{\langle k^2\rangle}{\langle k\rangle^2}$, where $\langle k\rangle$ denotes
the average degree~\cite{newman2010networks}.}
\begin{tabular}{c|c|c|c|c|c}
\hline
Network &N &M  &C &r &H\\
\hline
C.elegans &297 &2148  &0.308&-0.163&1.801\\ \hline
NS &1589 &2742 &0.791 &0.462 &2.011\\ \hline
FWEW &69 &880  &0.552 &-0.298 &1.275\\ \hline
FWFW &128 &2075  &0.335 &-0.112 &1.237 \\  \hline

USAir &332 &2126  &0.749 &-0.208 &3.464 \\ \hline
Jazz &198 &2742  &0.633 &0.02 &1.395 \\ \hline
Tap &1373 &6833  &0.557 &0.579 &1.644 \\ \hline
Power &4941 &6594  &0.107 &0.003 &1.45 \\ \hline

Metabolic &453 &2025&0.655 &-0.226 &4.485 \\ \hline
Yeast &2375 &11693  &0.388 &0.454 &3,476 \\ \hline
Router &5022 &6258  &0.033 &-0.138 &5.503 \\ \hline
PB &1222 &16724  &0.36 &-0.221 &2.971 \\ \hline
\end{tabular}\label{table1}
\end{table}

\section{Universality of PWCS phenomenon} \label{sec:universality}
\begin{figure}
\begin{center}
\includegraphics[width=3.0in]{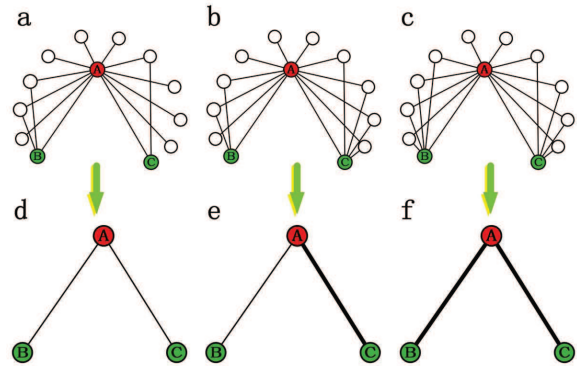}
\caption{(Color online) Degenerating the upper sketches into the lower cases by judging whether two links $\{B,A\}$ and $\{A,C\}$ are strong-tie link or common link.  Here we assume that if the number of common neighbors between A and B (or A and C) is larger than $\beta=3$, then the link is strong-tie link; otherwise, the link is common link in the opposite case. Fine lines and the bold lines in (d)-(f) are the common links and the strong-tie links, respectively. }
\label{fig1}
\end{center}
\end{figure}

To check whether the PWCS phenomenon commonly exists in real networks, we divide all links into common links or strong-tie links by judging whether the number of common neighbors between the two endpoints is larger than a threshold $\beta$. Taking Fig.~\ref{fig1} as an example, when we choose $\beta=3$, the links $\{A,B\}$ and $\{A,C\}$ in Figs.~\ref{fig1}(a), (b) and (c) can be correspondingly degenerated to the sketches in Fig~\ref{fig1}.~(d)-(f), where common links and  strong-tie links are marked by fine links and bold links, respectively.

\begin{figure}
\begin{center}
\includegraphics[width=2.5in]{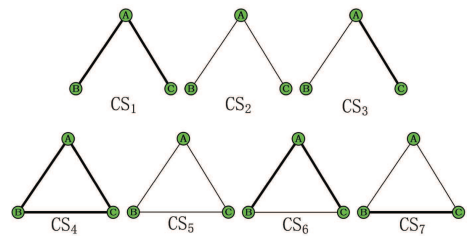}
\caption{(Color online) Seven possible configurations of connected subgraphs with three nodes. Fine lines and the bold lines are the common links and the strong-tie links, respectively.}
\label{fig2}
\end{center}
\end{figure}

In this paper, the threshold $\beta$ is chosen such that the number of common links and the number of strong-tie links are approximately equal in each network.
Once the value of $\beta$ is fixed, there are seven possible configurations for the connected subgraphs with 3 nodes (i.e., triples~\cite{newman2010networks} ), all the seven configurations are plotted in Fig.~\ref{fig2}, where the bold links and fine lines denote strong-tie links and common links, respectively. Let $N_i, i=1,\cdots,7$ be the number of $CS_i, i=1,\cdots,7$ (each CS represents a configuration in Fig.~\ref{fig2}) in networks. If both of $\{A,B\}$ and $\{A,C\}$ are strong-tie links, then the probability of node $B$ connecting node $C$ is defined as~\cite{lu2010link}:
\begin{eqnarray}\label{6}
P_1=\frac{3N_4+N_6}{N_1+3N_4+N_6}.
\end{eqnarray}
If only one of links $\{A,B\}$ or $\{A,C\}$ is strong-tie link, then the probability of node $B$ connecting node $C$ is defined as:
\begin{eqnarray}\label{7}
P_2=\frac{2N_6+2N_7}{2N_6+2N_7+N_3}.
\end{eqnarray}
If neither of them is strong-tie link, then the probability of node $B$ connecting node $C$ is:
\begin{eqnarray}\label{8}
P_3=\frac{3N_5+N_7}{3N_5+N_7+N_2}.
\end{eqnarray}

First, we define a subgraph with $n$ nodes be a weak clique the number of links among the $n$ nodes is rather dense, which is an extended definition of n-clique where all pairs of nodes are connected.
Next, by calculating the probability of node B connecting C, we can judge whether the phenomenon that nodes are preferentially linked to the nodes with weak clique structure (i.e., PWCS phenomenon) commonly exists in a network. We say that the PWCS phenomenon exists in the network if $P_1>P_2$ and $P_1>P_3$.  Moreover, we say that the PWCS phenomenon is significant if $P_1>P_2>P_3$, otherwise, the PWCS phenomenon is weak as  $P_1>P_3\geq P_2$.

Table~~\ref{table2} reports the values of $P_1$, $P_2$ and $P_3$ in the twelve real networks (labelled as RN) and the values on the corresponding null networks (labelled NN) are also comparatively shown. One can find that $P_1>P_2$ and $P_1>P_3$ in eleven networks except for Metabolic network ($P_1<P_3$, marked by red color). However, in the corresponding null networks, $P_1\approx P_2\approx P_3$. Also, for C.celegans, FWEW, FWFW, Power, Router and PB networks, where $P_1>P_2>P_3$. As a result, we can state that PWCS phenomenon is more significant in these six networks.

\begin{table}
\centering
\caption {The values of $P_1$, $P_2$ and $P_3$ in 12 real networks (RN) and the corresponding null networks (NN) are reported. Results in NN are marked in Italic.  Results in networks with significant PWCS, i.e., $P_1>P_2>P_3$ are shown in blue color, and results in Metabolic are marked by red color due to its specificity.}
\begin{tabular}{c|c|c|c|c}
\hline
Network                                              &Network       &$P_1$       &$P_2$          &\multicolumn{1}{|c}{$P_3$ } \\
\hline

\multicolumn{1}{c|}{\multirow {2}{*}{\textcolor[rgb]{0.00,0.07,1.00}{C.elegans}}}   &{\textcolor[rgb]{0.00,0.07,1.00}{RN}}       &\textcolor[rgb]{0.00,0.07,1.00}{0.2351}     & \textcolor[rgb]{0.00,0.07,1.00}{0.1654  }      &\multicolumn{1}{|c}{\textcolor[rgb]{0.00,0.07,1.00}{0.1519}}\\\cline{2-5}
\multicolumn{1}{c|}{}                               & {\emph{NN}}    &\emph{0.1011}     & \emph{0.0970}         &\multicolumn{1}{|c}{\emph{0.0953}} \\ \hline

\multicolumn{1}{c|}{\multirow {2}{*}{NS}}           &{RN}       &0.9292     & 0.2392        &\multicolumn{1}{|c}{0.5970}\\\cline{2-5}
\multicolumn{1}{c|}{}                               & {\emph{NN}}    &\emph{0 }         &\emph{ 0.0037 }       &\multicolumn{1}{|c}{\emph{0.0045}} \\ \hline

\multicolumn{1}{c|}{\multirow {2}{*}{\textcolor[rgb]{0.00,0.07,1.00}{FWEW}}}         &{\textcolor[rgb]{0.00,0.07,1.00}{RN}}       &\textcolor[rgb]{0.00,0.07,1.00}{0.5998}     & \textcolor[rgb]{0.00,0.07,1.00}{0.4832}        &\multicolumn{1}{|c}{\textcolor[rgb]{0.00,0.07,1.00}{0.2504}}\\\cline{2-5}
\multicolumn{1}{c|}{}                               & {\emph{NN}}    &\emph{0.7421}     & \emph{0.7627}         &\multicolumn{1}{|c}{\emph{0.7647}} \\ \hline

\multicolumn{1}{c|}{\multirow {2}{*}{\textcolor[rgb]{0.00,0.07,1.00}{FWFW}}}         &{\textcolor[rgb]{0.00,0.07,1.00}{RN}}       &\textcolor[rgb]{0.00,0.07,1.00}{0.4191}     & \textcolor[rgb]{0.00,0.07,1.00}{0.3532}        &\multicolumn{1}{|c}{\textcolor[rgb]{0.00,0.07,1.00}{0.1230}}\\\cline{2-5}
\multicolumn{1}{c|}{}                               & {\emph{NN}}    &\emph{0.5220 }    & \emph{0.5259 }        &\multicolumn{1}{|c}{\emph{0.5207}} \\ \hline

\multicolumn{1}{c|}{\multirow {2}{*}{USAir}}        &{RN}       &0.7008     & 0.1519        &\multicolumn{1}{|c}{0.2355}\\\cline{2-5}
\multicolumn{1}{c|}{}                               & {\emph{NN}}    &\emph{0.0752}     & \emph{0.0765 }        &\multicolumn{1}{|c}{\emph{0.0797}} \\ \hline

\multicolumn{1}{c|}{\multirow {2}{*}{Jazz}}         &{RN}       &0.6902     & 0.3968        &\multicolumn{1}{|c}{0.4503}\\\cline{2-5}
\multicolumn{1}{c|}{}                               & {\emph{NN}}    &\emph{0.2734 }    & \emph{0.2790}         &\multicolumn{1}{|c}{\emph{0.2804}} \\ \hline

\multicolumn{1}{c|}{\multirow {2}{*}{Tap}}          &{RN}       &0.7862     & 0.2969        &\multicolumn{1}{|c}{0.3673}\\\cline{2-5}
\multicolumn{1}{c|}{}                               & {\emph{NN}}    &\emph{0.0141 }    & \emph{0.0136 }        &\multicolumn{1}{|c}{\emph{0.0146}} \\ \hline

\multicolumn{1}{c|}{\multirow {2}{*}{\textcolor[rgb]{0.00,0.07,1.00}{Power}}}        &{\textcolor[rgb]{0.00,0.07,1.00}{RN}}       &\textcolor[rgb]{0.00,0.07,1.00}{0.2781}     & \textcolor[rgb]{0.00,0.07,1.00}{0.0854}        &\multicolumn{1}{|c}{\textcolor[rgb]{0.00,0.07,1.00}{0.0686}}\\\cline{2-5}
\multicolumn{1}{c|}{}                               & {\emph{NN}}    &\emph{0}          & \emph{0}             &\multicolumn{1}{|c}{\emph{0}} \\ \hline

\multicolumn{1}{c|}{\multirow {2}{*}{\textcolor[rgb]{1.00,0.00,0.00}{Metabolic}}}    &{\textcolor[rgb]{1.00,0.00,0.00}{RN}}       &\textcolor[rgb]{1.00,0.00,0.00}{0.1630}     &\textcolor[rgb]{1.00,0.00,0.00}{ 0.0760   }     &\multicolumn{1}{|c}{\textcolor[rgb]{1.00,0.00,0.00}{0.1643}}\\\cline{2-5}
\multicolumn{1}{c|}{}                               & {\emph{NN}}    &\emph{0.0409 }    & \emph{0.0374  }       &\multicolumn{1}{|c}{\emph{0.0389}} \\ \hline

\multicolumn{1}{c|}{\multirow {2}{*}{Yeast}}        &{RN}       &0.5945     & 0.1498        &\multicolumn{1}{|c}{0.1530}\\\cline{2-5}
\multicolumn{1}{c|}{}                               & {\emph{NN}}    &\emph{0.0089}     & \emph{0.0090  }       &\multicolumn{1}{|c}{\emph{0.0086}} \\ \hline

\multicolumn{1}{c|}{\multirow {2}{*}{\textcolor[rgb]{0.00,0.07,1.00}{Router}}}       &{\textcolor[rgb]{0.00,0.07,1.00}{RN}}       &\textcolor[rgb]{0.00,0.07,1.00}{0.1992}     & \textcolor[rgb]{0.00,0.07,1.00}{0.0254}        &\multicolumn{1}{|c}{\textcolor[rgb]{0.00,0.07,1.00}{0.0022}}\\\cline{2-5}
\multicolumn{1}{c|}{}                               & {\emph{NN}}    &\emph{0}          &\emph{ 0 }            &\multicolumn{1}{|c}{\emph{0} } \\ \hline

\multicolumn{1}{c|}{\multirow {2}{*}{\textcolor[rgb]{0.00,0.07,1.00}{PB}}}           &{\textcolor[rgb]{0.00,0.07,1.00}{RN}}       &\textcolor[rgb]{0.00,0.07,1.00}{0.3998}     & \textcolor[rgb]{0.00,0.07,1.00}{0.1247}        &\multicolumn{1}{|c}{\textcolor[rgb]{0.00,0.07,1.00}{0.0855}}\\\cline{2-5}
\multicolumn{1}{c|}{}                               & {\emph{NN}}    &\emph{0.0457}     & \emph{0.0454}         &\multicolumn{1}{|c}{\emph{0.0441}} \\ \hline

\end{tabular}\label{table2}
\end{table}

\section{Friend recommendation model} \label{sec:model}

\begin{figure}
\begin{center}
\includegraphics[width=2.5in]{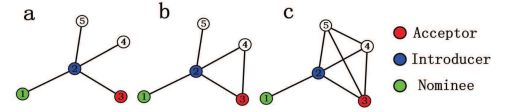}
\caption{(Color online) The role of PWCS on the probability of $f_{123}$. Node 2 (blue color, call introducer) want to introduce node 1 (green color, call nominee) to node 3 (red color, call acceptor).  The number common neighbor between node 2 and node 3 in (a), (b) and (c) is 0, 1 and 2, respectively. According to Eq.~(\ref{9}), one has (a) $f_{123}=1/3$; (b) $f_{123}=1/2$; (c) $f_{123}=1$. Namely, the probability of node 1 being introduced to node 3 in (c) is larger than (b) and is further larger than in (a). }
\label{fig3}
\end{center}
\end{figure}

Given that PWCS phenomenon commonly exists in real networks, whether can we design an effective link prediction method based on this phenomenon. Considering the cases in Fig.~\ref{fig3}, where node 3 ask its neighbor node 2 to introduce a friend to it. Since the number of common neighbors between node 2 and node 3 in Fig.~\ref{fig3}(c) is larger than that of in Fig.~\ref{fig3} (b) and is further larger than that of in Fig.~\ref{fig3}(a), in other words, the strength of link $\{2, 3\}$ in Fig.~\ref{fig3}(c) is the strongest. According to PWCS phenomenon, the probability (labelled by  $f_{123}$) of node 1 (call nominee, green color) being introduced to node 3 (call acceptor, red color) by node 2 (call introducer, blue color) in Fig.~\ref{fig3}(c) should be larger than that of in Fig.~\ref{fig3} (b), and then further larger than that of in Fig.~\ref{fig3}(a). To reflect the mentioned fact, we define $f_{ilj} $ be the probability of $i$ being introduced to $j$ by their common neighbor $l$, which is given as:
\begin{eqnarray}\label{9}
f_{ilj}=\frac{1}{k(l)-1-|\Gamma(l)\cap\Gamma(j)|}.
\end{eqnarray}
Based on the definition in Eq.~(\ref{9}), the values of $f_{123}$ in Fig.~\ref{fig3}(a), (b) and (c) are 1/3, 1/2 and 1, respectively. That is to say, the probability $f_{ilj}$ can reflect the PWCS phenomenon in real networks.

More importantly, Eq.~(\ref{9}) addresses two important facts: first, since node $l$ will not introduce node $j$ to $j$, as a result, 1 is subtracted in denominator of Eq.~(\ref{9}); second, in social communication, when a friend introduce one of his friends to me, he should introduce his friends but \emph{excluding} the common friends. Therefore, the common neighbors set between $j$ and $l$ (i.e., $\Gamma(l)\cap\Gamma(j)$) should be subtracted in denominator of Eq.~(\ref{9}). For instance, in  Fig.~\ref{fig3}(c), node 2 will not introduce node 3 to node 3, and nodes 4 and 5 should not be introduced to node 3.

Let $f_{ij}$ be the weight of node $i$ being introduced to node $j$ (we use weight rather than  probability since $f_{ij}$ may larger than 1), which is written as:
\begin{eqnarray}\label{10}
f_{ij}=\sum_{l\in\Gamma(i)\cap\Gamma(j)}f_{ilj}=\sum_{l\in\Gamma(i)\cap\Gamma(j)}\frac{1}{k(l)-1-S^{CN}_{jl}}.
\end{eqnarray}
Here the value of $f_{ij}$ increases with the number of common neighbors.

With the above preparations, the similarity index $S^{FR}_{ij}$ for a pair of nodes $i$ and $j$ is defined as
\begin{eqnarray}\label{11}
S^{FR}_{ij}=\frac{f_{ij}+f_{ji}}{2},
\end{eqnarray}
which guarantees $S^{FR}_{ij}=S^{FR}_{ji}$.

\begin{figure}
\begin{center}
\includegraphics[width=3.0in]{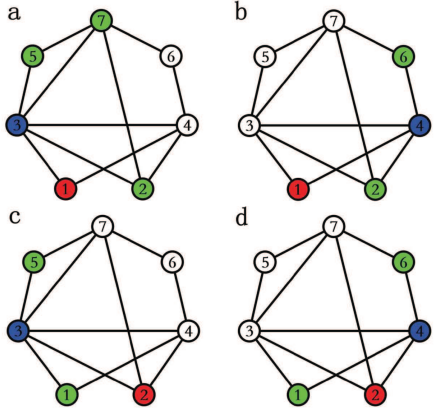}
\caption{(Color online) Calculation of the similarity $S^{FR}_{12}$ between node 1 and node 2. Nodes 1 and 2 can be introduced by their common neighbors 3 and 4.  (a) node 3 introduces his friends to node 1 . Only neighbor nodes 2, 5, 7 can be introduced to node 1 but excludes node 4, since node 4 has been a friend of node 1. Thus,  the probability of node 3 introducing node 2 to node 1 is: $f_{231}=1/3$; (b) node 2 is introduced to node 1 by node 4, here only nodes 2 and 6 can be introduced to node 1. As a result, the probability $f_{241}=1/2$; (c) node 1 is introduced to node 2 by node 3, here only nodes 1 and 5 can be introduced to node 1. As a result, the probability $f_{132}=1/2$; (d) node 1 is introduced to node 2 by node 4, here only nodes 1 and 6 can be introduced to node 1. As a result, the probability $f_{141}=1/2$. We have $f_{21}=1/3+1/2$ by combing (a) and (b), and $f_{12}=1/2+1/2$ by combing (c) and (d). So the FR similarity index is $S^{FR}_{12}=S^{FR}_{21}=11/12$. }
\label{fig4}
\end{center}
\end{figure}

The sketches in Fig.~\ref{fig4} is given to show how to calculate the similarity between node 1 and node 2 based on the FR index. Also, the red, blue and green nodes denote the acceptors, introducers and nominees, respectively. Node 2 can be introduced to node 1 by node 3 (see Fig.~\ref{fig4}(a)) or node 4 (see Fig.~\ref{fig4}(b)). When node 3 is an introducer (see Fig.~\ref{fig4}(a)), who will introduce nodes 2, 5 and 7 (green color) to node 1 with equal probability, \emph{but excludes node 4}, i.e., $f_{231}=1/3$. Similarly, when node 4 is an introducer (see Fig.~1(b)), who just introduces nodes 2 and 6 (green color) to node 1 with equal probability, \emph{but excludes node 3}, i.e., $f_{241}=1/2$. Therefore, the probability $f_{21}=1/3+1/2=5/6$. Likely, from Figs.~\ref{fig5}(c) and (d), the value of $f_{12}=1/2+1/2=1$. Therefore, the FR similarity index is $S^{FR}_{12}=S^{FR}_{21}=11/12$.

Combing Eqs.~(\ref{9}), (\ref{10}) and (\ref{11}), the advantages of FR index can be summarized as: 1) similar to many local similarity indices, the similarity between a pair of nodes increases with the number of common neighbors; 2) like AA index and RA index, FR index depresses the contribution of the high-degree common neighbors; 3) most importantly, FR index can make use of the PWCS phenomenon in many real networks; 4) FR index has higher resolution than other local similarity indices. For instance, the similarities $S^{CN}_{13}$, $S^{AA}_{13}$ and $S^{RA}_{13}$ are the same in Figs.~\ref{fig3}(a), (b) and (c). Yet, the value of $S^{FR}_{13}$ in Fig.~\ref{fig3}(c) is larger than Fig.~\ref{fig3}(b), and is further larger than Fig.~\ref{fig3}(a).

%
%

\section{Results} \label{sec:main results}
In this section, the comparison of FR index with CN, AA and RA indices in twelve networks is summarized in Tab.~~\ref{table3}. As shown in Tab.~~\ref{table3}, FR index in general outperforms the other three indices in link prediction, regardless of AUC or Precision. The highest accuracy in each line is emphasized in bold.

\begin{table}
\centering
\caption {Comparison of $S^{FR}$ with $S^{CN}$, $S^{AA}$ and $S^{RA}$ in 12 networks, including AUC and Precision. The highest value in each row is marked in bold.}
\begin{tabular}{c|c|c|c|c|c|c}
\hline
Network                                            &Metric             &CN         &AA            &RA           &\multicolumn{1}{|c}{FR} \\
\hline

\multicolumn{1}{c|}{\multirow {2}{*}{C.elegans}}   &{AUC}           &0.8501      & 0.8663      &0.8701          &\multicolumn{1}{|c}{\textbf{0.8756}}\\
\cline{2-6}

\multicolumn{1}{c|}{}                              & {Precision}     &0.1306     & 0.1374      & 0.1315          &\multicolumn{1}{|c}{\textbf{0.1504}} \\
\hline

\multicolumn{1}{c|}{\multirow {2}{*}{NS}}          &{AUC}           &0.9913     & 0.9916    &\textbf{0.9917}     &\multicolumn{1}{|c}{0.9916}\\
\cline{2-6}

\multicolumn{1}{c|}{}                              & {Precision}     &0.8707    & 0.9731      & 0.9712           &\multicolumn{1}{|c}{\textbf{0.9832}} \\
\hline

\multicolumn{1}{c|}{\multirow {2}{*}{FWEW}}        &{AUC}           &0.6868     & 0.6939        & 0.7017         &\multicolumn{1}{|c}{\textbf{0.7595}}\\
\cline{2-6}

\multicolumn{1}{c|}{}                              & {Precision}     &0.1415    & 0.1551      & 0.1664           &\multicolumn{1}{|c}{\textbf{0.2763}} \\
\hline

\multicolumn{1}{c|}{\multirow {2}{*}{FWFW}}        &{AUC}           &0.6074    & 0.6097       & 0.6142         &\multicolumn{1}{|c}{\textbf{0.6623}}\\
\cline{2-6}

\multicolumn{1}{c|}{}                              & {Precision}     &0.0837    & 0.0853      & 0.082           &\multicolumn{1}{|c}{\textbf{0.1798}} \\
\hline

\multicolumn{1}{c|}{\multirow {2}{*}{USAir}}       &{AUC}           &0.9558    & 0.9676       & 0.9736         &\multicolumn{1}{|c}{\textbf{0.9752}}\\
\cline{2-6}

\multicolumn{1}{c|}{}                              & {Precision}     &0.606    & 0.6218      & 0.6337           &\multicolumn{1}{|c}{\textbf{0.6586}} \\
\hline

\multicolumn{1}{c|}{\multirow {2}{*}{Jazz}}        &{AUC}           &0.9563     & 0.963    &\textbf{ 0.9717}         &\multicolumn{1}{|c}{0.9714}\\
\cline{2-6}

\multicolumn{1}{c|}{}                              & {Precision}     &0.8247    & 0.8401      & 0.8192           &\multicolumn{1}{|c}{\textbf{0.8406}} \\
\hline

\multicolumn{1}{c|}{\multirow {2}{*}{Tap}}         &{AUC}           &0.9538    & 0.9545       & 0.9548         &\multicolumn{1}{|c}{\textbf{0.955}}\\
\cline{2-6}

\multicolumn{1}{c|}{}                              & {Precision}     &0.7594    & 0.78      & 0.7818           &\multicolumn{1}{|c}{\textbf{0.8659}} \\
\hline

\multicolumn{1}{c|}{\multirow {2}{*}{Power}}       &{AUC}           &0.6249    &\textbf{0.6251}  & 0.6245         &\multicolumn{1}{|c}{0.6248}\\
\cline{2-6}

\multicolumn{1}{c|}{}                              & {Precision}     &0.1215    & 0.0952      & 0.0801           &\multicolumn{1}{|c}{\textbf{0.1275}} \\
\hline

\multicolumn{1}{c|}{\multirow {2}{*}{Metabolic}}    &{AUC}           &0.9248    & 0.9565       & 0.9612        &\multicolumn{1}{|c}{\textbf{0.9623}}\\
\cline{2-6}

\multicolumn{1}{c|}{}                              & {Precision}     &0.2026    & 0.2579      & 0.3219           &\multicolumn{1}{|c}{\textbf{0.3302}} \\
\hline

\multicolumn{1}{c|}{\multirow {2}{*}{Yeast}}       &{AUC}           &0.9158    & 0.9161       & 0.9167        &\multicolumn{1}{|c}{\textbf{0.9172}}\\
\cline{2-6}

\multicolumn{1}{c|}{}                              & {Precision}     &0.6821    & 0.6958      & 0.4988         &\multicolumn{1}{|c}{\textbf{0.8041}} \\
\hline

\multicolumn{1}{c|}{\multirow {2}{*}{Router}}       &{AUC}           &0.6519    &\textbf{0.6523}  & 0.652         &\multicolumn{1}{|c}{0.6519}\\
\cline{2-6}

\multicolumn{1}{c|}{}                              & {Precision}     &\textbf{0.1144}    & 0.1104      & 0.0881    &\multicolumn{1}{|c}{0.0592} \\
\hline

\multicolumn{1}{c|}{\multirow {2}{*}{PB}}         &{AUC}              &0.9239            &0.9275     & 0.9286      &\multicolumn{1}{|c}{\textbf{0.9309}}\\
\cline{2-6}

\multicolumn{1}{c|}{}                              & {Precision}     &\textbf{0.4205}    & 0.3782     & 0.2509           &\multicolumn{1}{|c}{0.3454} \\
\hline
\end{tabular}\label{table3}
\end{table}

\begin{figure*}
\begin{center}
\includegraphics[width=6.0in]{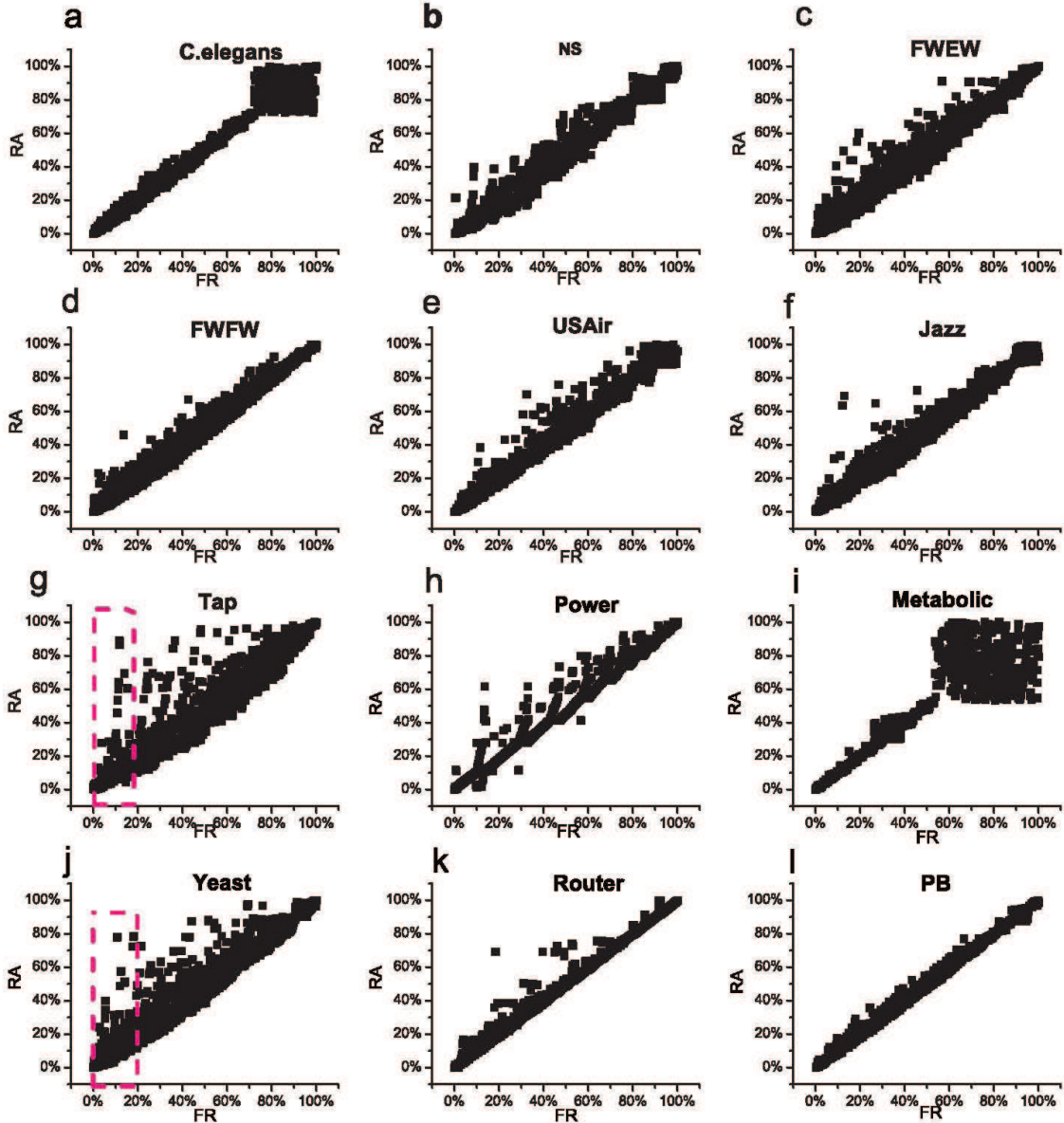}
\caption{(Color online) The correlation of ranking values between FR index and RA index based on Precision.  The percentage values in x-axis and y-axis are the top percentage ranking values of FR index and RA index, respectively.  The regions marked by pink dash boundary in subfigures (g) and (j) correspond to the cases in which some links have higher FR ranking values but have low RA ranking values. }
\label{fig5}
\end{center}
\end{figure*}

Moreover, the correlation of ranking values between FR index and RA index is given in Fig.~\ref{fig5}, where the percentage values in x or y axis is the top percentage of ranking values based on Precision. As a result, a small percentage value means a higher ranking value. Fig.~\ref{fig5} describes that a high RA ranking value of links gives rise to a high FR ranking value. However, a high FR ranking value of links may induce a low RA ranking value of links. Take Tap and Yeast networks as examples, based on FR index, some links have higher ranking values, however their corresponding ranking values based on RA index may very small (see the regions marked by pink dash boundary in Figs.~\ref{fig5}(g) and (j)).

\begin{figure}
\begin{center}
\includegraphics[width=3.0in]{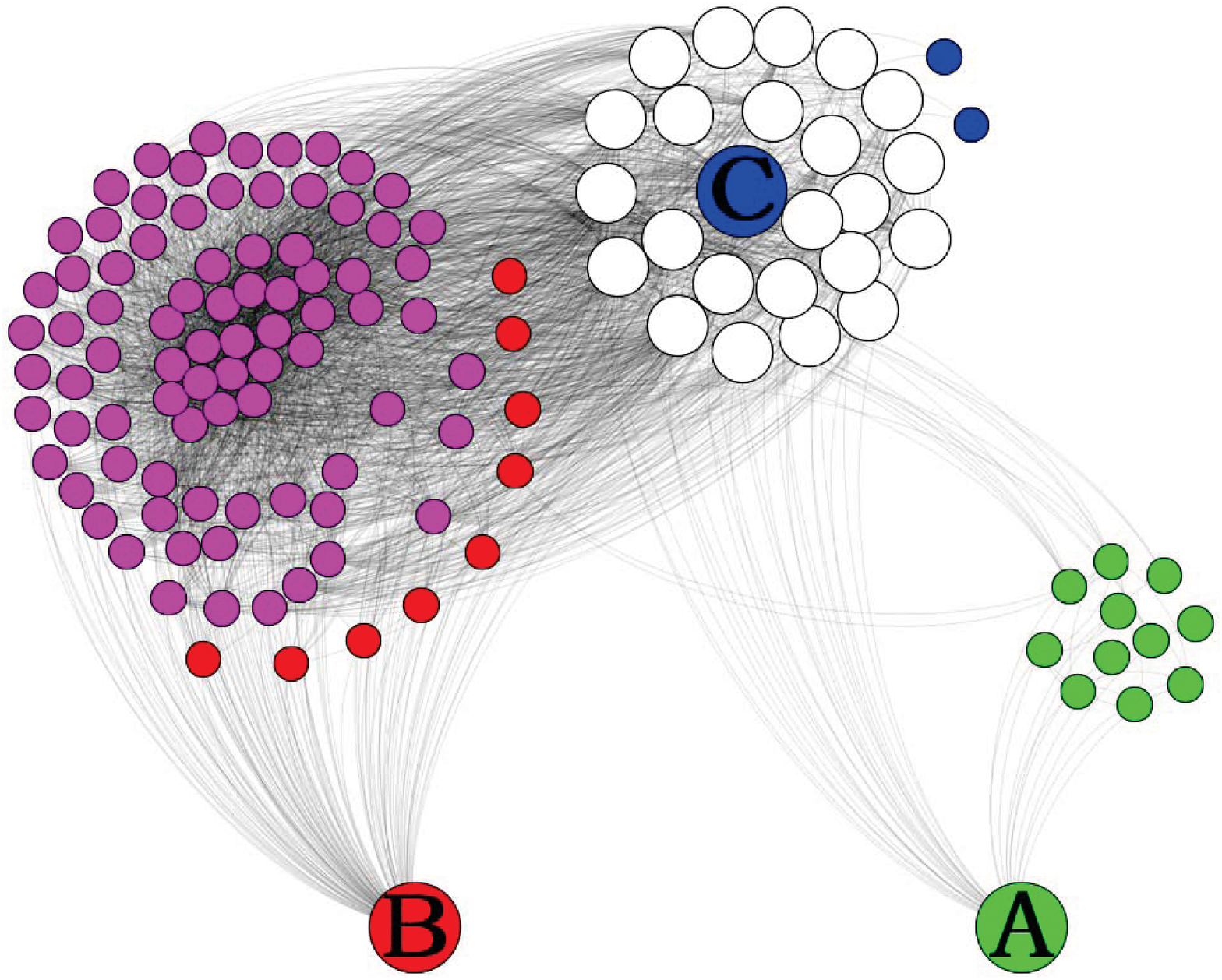}
\caption{(Color online) A typical case in the Yeast network is considered to emphasize the difference between FR index and RA index, where nodes A, B and C are the node 1175, 421 and 205 in the Yeast network. Two links $\{A,C\}$ and $\{B,C\}$ share a common endpoint C, and both of them are strong-tie links. Therefore, the similarity $S^{FR}_{AB}$ is rather large. However, when using RA index, the ranking number of $S^{RA}_{AB}$ is very low owing to the large degree value of node C, causing the failure of RA index in predicting such an existing link. Red nodes, green nodes  and blue nodes are the neighbors of A, B and C (including themselves),  respectively.  Purple nodes are the common neighbors of A and C; white nodes are the common neighbors of A, B and C.}
\label{fig6}
\end{center}
\end{figure}

By analyzing a typical case in the Yeast network (see Fig.~\ref{fig6}), where two nodes A and B are the neighbors of introducer C (in fact, there has a link connecting A and B in the Yeast network). Since links $\{A,C\}$ and $\{B,C\}$ are strong-tie links. When using FR index, the similarity $S^{FR}_{AB}$ is rather large, which can predict the existence of link $\{A,B\}$. However, for RA index, since the large degree value of introducer C, the similarity $S^{RA}_{AB}$ is very small, such an existing link $\{A,B\}$ cannot be accurately predicted by RA index.

\section{Role of PWCS} \label{sec:PWCS}

We have validated that the FR index based on PWCS phenomenon can improve the performance of link prediction, and the reasons were also analyzed. Here we want to know how the strength of PWCS affects the performance of RA index and FR index. For this purpose, we propose a generalized friend recommendation (GFR) index, which is given as:

\begin{eqnarray}\label{12}
S^{GFR}_{ij}=\frac{1}{2}\sum_{l\in\Gamma(i)\cap\Gamma(j)}\big(\frac{1}{k(l)-\alpha S^{CN}_{jl}}+\frac{1}{k(l)-\alpha S^{CN}_{il}}\big),
\end{eqnarray}
where parameter $0\leq\alpha\leq 1$ is used to uncover the role of PWCS. As $\alpha=0$, Eq.~(\ref{12}) returns to RA index, that is, $S^{GFR}_{ij}=S^{RA}_{ij}$. When $\alpha=1$, the difference between FR method and GFR method is the absence of 1 in the denominators of Eq.~(\ref{12}), therefore, we can simply view GFR index is the same as FR index when $\alpha=1$. As a result, with the increasing of $\alpha$ from zero to one, $S^{SFR}$ index can comprehensively investigate the role of PWCS on the RA index and FR index.

\begin{figure*}
\begin{center}
\includegraphics[width=6.0in]{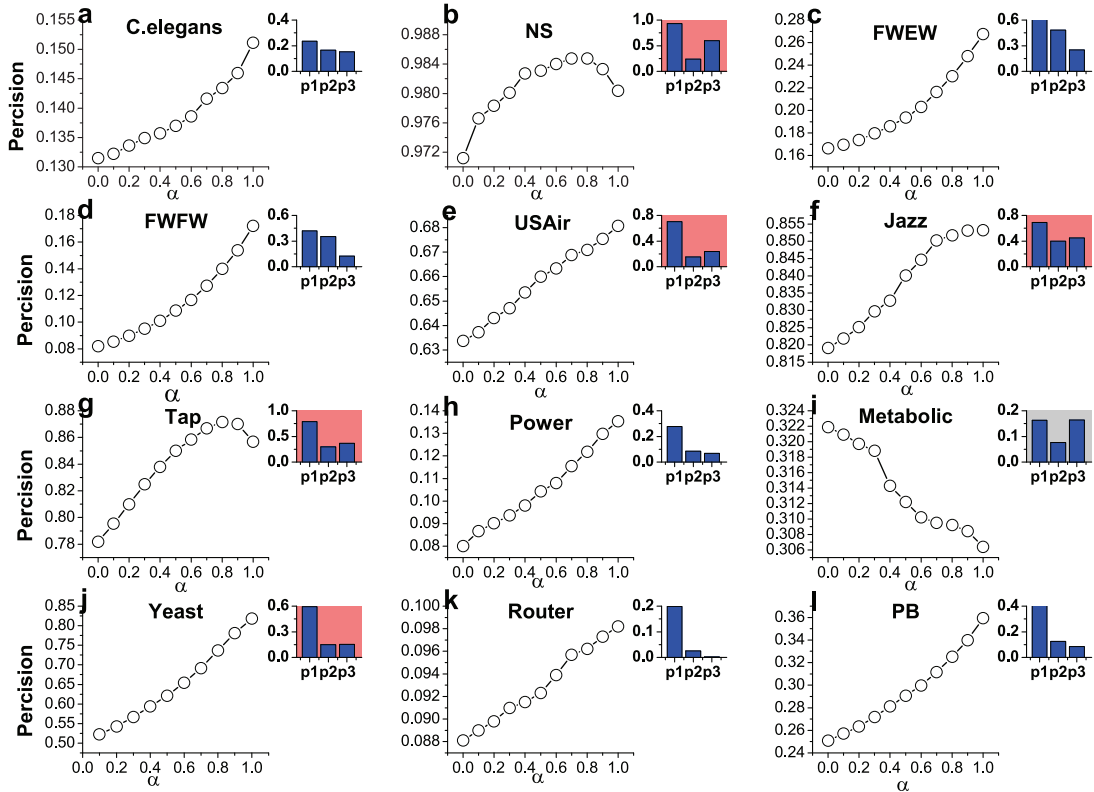}
\caption{(Color online) Effects of $\alpha$ in Eq.~(\ref{13}) on Precision are plotted in 12 networks. Inset in each subfigure is to show the values of $P_1$, $P_2$ and $P_3$. The background of inset is white color when $P_1>P_2>P_3$; the background of inset is light red color when $P_1>P_3\geq P_2$. Otherwise, the background of inset for Metabolic is gray color. }
\label{fig7}
\end{center}
\end{figure*}

The effect of $\alpha$ on the Precision in all twelve networks is plotted in Fig.~\ref{fig7}. As illustrated in Fig.~\ref{fig7}, several interesting phenomena and meaningful conclusions can be summarized: First, except for Metabolic network, the Precision for the case of $\alpha>0$ is far larger than the case of $\alpha=0$ (i.e., RA index) in all other 11 networks. Since $P_1>P_2$ and $P_1>P_3$ in these 11 networks, which indicates that PWCS phenomenon in networks can ensure the higher accuracy of FR index (i.e.,$\alpha=1$) in link prediction ; Second, Metabolic network has \emph{non-PWCS phenomenon} since $P_1>P_2$ and $P_2<P_3$, and Fig.~\ref{fig7}(i) suggests that Precision \emph{decreases} with the value of $\alpha$. In other words, FR index is invalid in network with non-PWCS phenomenon, which again emphasizes the importance of PWCS in link prediction; At last, by systematically comparing the subfigures in Fig.~\ref{fig7}, one can see that, when the networks with \emph{weak PWCS} $P_1>P_3\geq P_2$ (i.e., the insets are light red background, see Fig.~\ref{fig7}(b), (e), (f), (g) and (j)), Precision increases with $\alpha$ at first and then decreases when $\alpha$ is further increased (except Fig.~\ref{fig7}(e)). However, when $P_1>P_2>P_3$ (i.e., networks with \emph{significant PWCS}, the insets are white background, see Figs.~\ref{fig7}(a), (c), (d), (h), (k) and (l)), Precision \emph{always} increases with the value of $\alpha$ even when $\alpha=1.0$.

In view of this observation, we can conjecture the role of PWCS can be further explored when the PWCS phenomenon is significant. Unfortunately, the maximal value $\alpha$ in Eq.~(\ref{12}) is one. the denominator may be negative if $\alpha>1$. So we design a new index to further explore the role of significant PWCS.

Since Eq.~(\ref{12}) can be rewritten as
\begin{eqnarray}\label{13}
S^{GFR}_{ij}=\frac{1}{2}\sum_{l\in\Gamma(i)\cap\Gamma(j)}\frac{2k(l)-S^{CN}_{il}-S^{CN}_{jl}}{(k(l)-S^{CN}_{jl})(k(l)-\alpha S^{CN}_{il})}
\end{eqnarray}
when $\alpha=1$. To further play the role of PWCS, another similarity index, called strong friend recommendation (labelled as SFR) index, is given in following
\begin{eqnarray}\label{14}
S^{SFG}_{ij}=\frac{1}{2}\sum_{l\in\Gamma(i)\cap\Gamma(j)}\frac{2k(l)}{(k(l)-S^{CN}_{jl})(k(l)-\alpha S^{CN}_{il})}.
\end{eqnarray}
Combing Eq.~(\ref{13}) with Eq.~(\ref{14}), we can find that two subtrahends $S^{CN}_{il}$ and $S^{CN}_{jl}$ in the numerator of Eq.~(\ref{13}) are removed. So Eq.~(\ref{14}) can better play the role of PWCS.

We conjecture that the performance of SFR index is better than GFR index when $P_1>P_2>P_3$ (i.e., significant PWCS), and worse than that of GFR index when $P_1<P_2$ and $P_1<P_3$ (i.e., non-PWCS). However, it is difficult to distinguish which one has better performance when $P_1>P_3\geq P_2$(i.e., weak PWCS). As presented in Tab.~~\ref{table4}, Precision in 12 networks validates our conjecture.

\begin{table*}
\centering
\caption {The comparison of Precision between SFR index and GFR index ($\alpha=1$) in 12 networks. The results suggest that the accuracy of link prediction can be further improved by SFR index when the networks have significant PWCS.  The highest value in each case is marked as bold. }
\begin{tabular}{c|c|c|c|c|c|c}
 \hline
\multicolumn{1}{c|}{ }                                       &\multicolumn{6}{|c}{ $P_1>P_2>P_3$ }\\
\hline

Index                   &C.elegans           &FWEW               &FWFW            &Power              &Router            &PB\\ \hline
GFR ($\alpha$=1)        &0.1511               &0.2676             &0.172           &0.1354             &0.0982            &0.3595\\ \hline
SFR  ($\alpha$=1)                    &\textbf{0.1577}      &\textbf{0.2912}    &\textbf{0.2057} &\textbf{0.1658}    &\textbf{0.112}    &\textbf{0.4353} \\\hline

\multicolumn{1}{c|}{ }               &\multicolumn{5}{|c}{ $P_1>P_3\geq P_2$ }                          &\multicolumn{1}{|c}{ $P_1<P_3,P_2<P_3$ }\\
\hline
Index                   &NS                   &USAir              &Jazz             &Tap                &Yeast             &Metabolic\\ \hline
GFR ($\alpha$=1)        &\textbf{0.9804}      &0.6807             &0.8532           &\textbf{0.8568}    &0.8178            &\textbf{0.3064}\\ \hline
SFR     ($\alpha$=1)                 &0.9744               &\textbf{0.6866}    &\textbf{0.8739}  &0.8485             &\textbf{0.8587}   &0.2912\\\hline

\end{tabular}\label{table4}
\end{table*}

Synthesizing the above results, we can find that the ranking of $P_1$, $P_2$ and $P_3$ has determinant effect on the performance of the proposed index. Inspired by this clue, we may design a universal indicator to do link prediction based on the values of $P_1$, $P_2$ and $P_3$ in different networks. To this end, we design a mixed friend recommendation (labelled MFR) index:

\begin{numcases}{S^{MFR}_{ij}=}\label{15}
\nonumber S^{SFR}_{ij},~~~~~~~~~~~~~ \mbox{$P_1>P_2>P_3$};\\
\nonumber S^{FR}_{ij} ,~~~~~~~~~~~~\mbox{$P_1>P_2$, $P_1>P_3$, $P_2<P_3$};\\
 S^{RA}_{ij},~~~~~~~~~~~~~~~\mbox {otherwise}.
\end{numcases}

Table~5 lists the results of MFR index and FR index in 7 networks (since MFR index is the same to FR index when $P_1>P_2$, $P_1>P_3$ and $P_2<P_3$, in this case, it is unnecessary to compare the two indices). The results in Tab.~~\ref{table5} indicate that, compared with FR index, MFR index can further improve the accuracy of link prediction.

\begin{table*}
\centering
\caption {Comparison of Precision between FR index and WFR index in 7 networks. The highest value in each case is given in bold.}
\begin{tabular}{c|c|c|c|c|c|c|c|c}
\hline
  Metric   &Index      &C.elegans      &FWEW             &FWFW           &Power           &Router          &PB              &Metabolic \\\hline
\multicolumn{1}{c|}{\multirow {2}{*}{AUC}}
           &{FR}       &0.8756          &0.7595           &0.6623         &\textbf{0.6248} &\textbf{0.6519} &0.9309          &\textbf{0.9623}\\\cline{2-9}

\multicolumn{1}{c|}{}
          & {WFR}      &\textbf{0.8771} &\textbf{0.7771} &\textbf{0.6878} &0.6247          &0.6516          &\textbf{0.9314} &0.9612 \\\hline

\multicolumn{1}{c|}{\multirow {2}{*}{Precision}}
           &{FR}       &0.1504          &0.2763           &0.1798         &0.1275          &0.0592          &0.3454         &\textbf{0.3302}\\\cline{2-9}

\multicolumn{1}{c|}{}
          & {WFR}      &\textbf{0.1577} &\textbf{0.2912} &\textbf{0.2057} &\textbf{0.1658} &\textbf{0.112}  &\textbf{0.4353}&0.3219 \\\hline

\end{tabular}\label{table5}
\end{table*}

\section{Conclusions} \label{sec:discussion}

In summary, by analyzing the structural properties in real networks, we have found that there exists a common phenomenon: nodes are preferentially linked to the nodes with weak clique structure. Then we have proposed a friend recommendation model to better predict the missing links based on the significant phenomenon. Through the detailed analysis and experimental results, we have shown that FR index has several typical characteristics: First, FR index is based on the information of common neighbors, which is a local similarity index. Thus, the algorithm is simple and has low complexity; Second, the common neighbors with small degrees has greater contributions than the common neighbors with larger degrees; Third, FR index can take full advantage of the PWCS phenomenon, and so forth.

Furthermore, we also proposed an SFR index to further improve the accuracy of link prediction when networks have \emph{significant} PWCS phenomenon.  At last, by judging whether the networks have significant PWCS, weak PWCS or non-PWCS phenomenon, we have also proposed a mixed friend recommendation index which can increase the accuracy of link prediction in different networks. In this work, we mainly applied FR index to unweighed and undirected networks, and how to generalize our FR index to weighted~\cite{zhao2015prediction,aicher2015learning} or directed networks~\cite{guo2013predicting} is our further purpose.

\begin{acknowledgments}
This work is supported by the National Natural Science Foundation of China (Grant Nos.~ 61473001), and partially supported by open fund of Key Laboratory of Computer Network and Information Integration (Southeast University), Ministry of Education (No. K93-9-2015-03B).
\end{acknowledgments}

\end{document}